# Absolute Quantification of *sp*$^3$ Defects in Semiconducting Single-Wall Carbon Nanotubes by Raman Spectroscopy

Finn L. Sebastian[1], Nicolas F. Zorn[1], Simon Settele[1], Sebastian Lindenthal[1], Felix J. Berger[1,§], Christoph Bendel[1], Han Li[2], Benjamin S. Flavel[2] and Jana Zaumseil[1,*]

[1]Institute for Physical Chemistry, Universität Heidelberg, D-69120 Heidelberg, Germany

[2]Institute of Nanotechnology, Karlsruhe Institute of Technology, D-76131 Karlsruhe, Germany

**Corresponding Author**

*E-mail: zaumseil@uni-heidelberg.de

§**Present address:**

Center for Nano Science and Technology @PoliMi, Istituto Italiano di Tecnologia, I-20133 Milano, Italy

# Abstract

The functionalization of semiconducting single-wall carbon nanotubes (SWCNTs) with luminescent $sp^3$ defects creates red-shifted emission features in the near-infrared and boosts their photoluminescence quantum yields (PLQYs). While multiple synthetic routes for the selective introduction of $sp^3$ defects have been developed, a convenient metric to precisely quantify the number of defects on a SWCNT lattice is not available. Here, we present a direct and simple quantification protocol based on a linear correlation of the integrated Raman D/G$^+$ signal ratios and defect densities as extracted from PLQY measurements. Corroborated by a statistical analysis of single-nanotube emission spectra at cryogenic temperature, this method enables the quantitative evaluation of $sp^3$ defect densities in (6,5) SWCNTs with an error of ± 3 defects per µm and the determination of oscillator strengths for different defect types. The developed protocol requires only standard Raman spectroscopy and is independent of the defect configuration, dispersion solvent and nanotube length.

The chemical modification of semiconducting single-wall carbon nanotubes (SWCNTs) is a versatile tool to tune their properties for various applications such as quantum-light sources,[1-3] sensing or bioimaging.[4-7] In particular, covalent functionalization of SWCNTs with luminescent $sp^3$ defects (also referred to as organic color centers) creates red-shifted emissive states in the near-infrared (NIR).[8-12] These states exhibit deep optical trap potentials (100 - 250 meV), which are able to localize the highly mobile excitons[9] that would otherwise explore large nanotube segments to encounter quenching sites or decay radiatively by $E_{11}$ emission (see **Figure 1a**).[13] By preventing excitons from reaching quenching sites and decaying non-radiatively, these $sp^3$ defects can increase the total photoluminescence quantum yield (PLQY) of SWCNTs.[9, 14] However, precise control of the degree of $sp^3$ functionalization is crucial. The maximum ensemble PLQY is observed at fairly low levels of functionalization, which would be favorable for electrically pumped light-emitting devices.[15] In contrast, single-photon emission requires exactly one luminescent defect per SWCNT.[1] Hence, accurate control over the degree of functionalization and knowledge of the precise $sp^3$ defect densities are highly desired for further optimization.

The emission wavelengths of functionalized SWCNTs are predominantly determined by the binding configuration of the defects, as two $sp^2$ carbon atoms must be converted to $sp^3$ carbons to form one defect state. In chiral SWCNTs there are six possible relative positions of the involved carbon atoms, all of which lead to different optical trap depths and photoluminescence (PL) peak wavelengths.[16-17] However, only two of them are commonly found in functionalized (6,5) SWCNTs and give rise to separate NIR emission peaks termed $E_{11}$* and $E_{11}$*$^-$, the latter being more red-shifted and exhibiting a longer fluorescence lifetime than the former.[5, 18]

Various synthetic methods have been developed in an attempt to control the degree and type of $sp^3$ functionalization of nanotube dispersions in water or organic solvents.[14, 18-20] But, comparing different reports on functionalized SWCNTs and their properties is difficult due to the use of indirect metrics for their quantification. For example, PL peak intensity ratios strongly depend on the specific experimental setups and excitation power. They only provide a relative but not an absolute defect density.[21] Currently, the most reliable technique to determine the number of luminescent defects on individual SWCNTs is to count distinct emission peaks at cryogenic temperatures.[5, 22] However, this method requires substantial experimental effort and tedious statistical analysis.

For graphene, a simple approach to quantify lattice defect densities using Raman spectroscopy is already well-established.[23-24] The introduction of point-like defects into the planar $sp^2$ carbon lattice leads to the activation of the Raman D mode. Its relative intensity compared to the G mode can be used as a direct metric for the areal defect density.[25] The corresponding equation has been applied as means of quality control in graphene samples[26] and to monitor the degree of chemical functionalization.[27] It was recently modified and extended to include line defects in graphene.[28] Although the Raman D mode of SWCNTs also reflects the degree of structural disorder and number of defects,[29-32] no quantitative relation to the absolute density of defects, especially at low defect densities, has been reported so far.

Here, we present a robust empirical metric for the absolute quantification of $sp^3$ defects in the most commonly used (6,5) SWCNTs. Our method is based on a cross-correlation of Raman spectra, PLQY data and statistics of low-temperature single-nanotube PL measurements. The final protocol only requires resonant Raman spectroscopy of drop-cast SWCNT films. That way, we are able to extract the absolute $sp^3$ defect density, independent of type, within an error of ± 3 defects per micrometer.

To produce nanotube samples with a controlled number of defects per nanotube length, polymer-sorted (6,5) SWCNTs were functionalized using two different procedures (see **Methods**, **Supporting Information**). $E_{11}$* defects (emission at ~1170 nm) were introduced by treatment with 4-nitrobenzenediazonium tetrafluoroborate ($DzNO_2$) in a mixture of toluene/acetonitrile employing a phase-transfer agent.[14] More red-shifted $E_{11}^{*-}$ defects (emission at ~1250 nm) were created by reaction with 2-iodoaniline in the presence of the organic base potassium *tert*-butoxide (KO$^t$Bu).[18] The degree of $sp^3$ functionalization was controlled by variation of the DzNO$_2$ concentration or by adjusting the reaction time with 2-iodoaniline. **Figures 1b** and **1c** show the corresponding normalized PL spectra of selectively functionalized (6,5) SWCNT dispersions collected under pulsed excitation at the $E_{22}$ transition (575 nm). Furthermore, we employed a sequential reaction scheme to create (6,5) SWCNTs with controlled concentrations of $E_{11}^*$ and $E_{11}^{*-}$ defects as shown in **Figure 1d**.

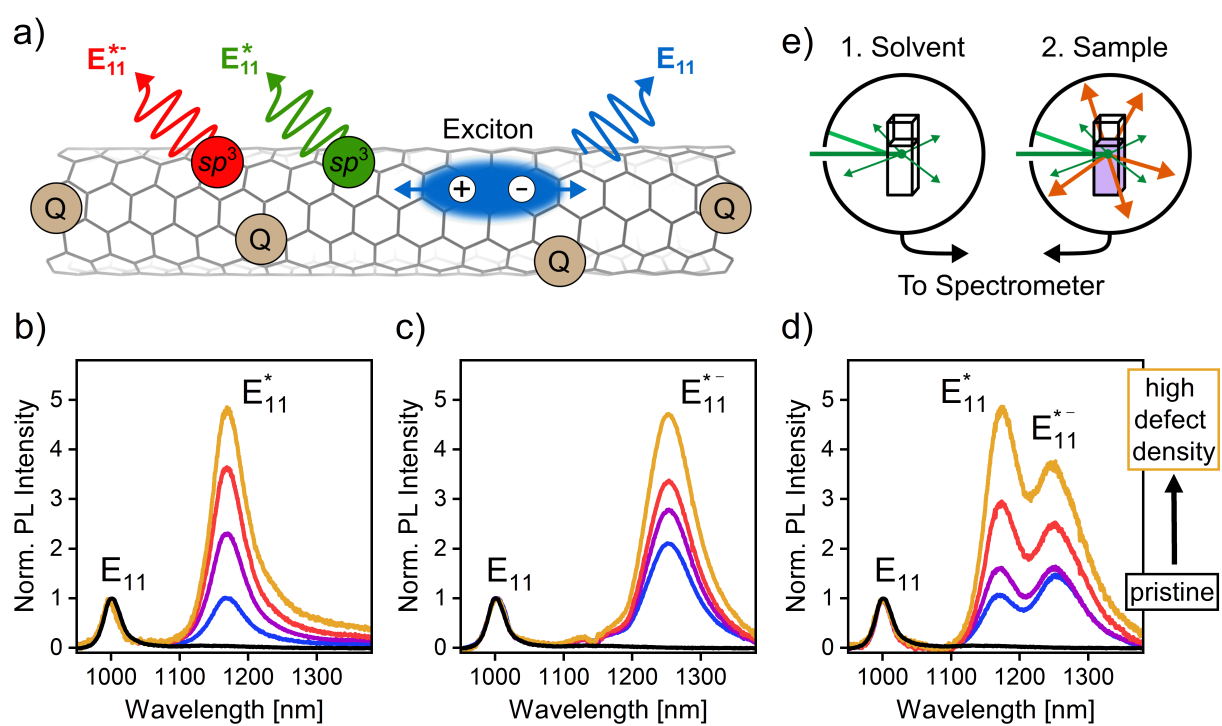

**Figure 1. (a)** Schematic illustration of a (6,5) SWCNT functionalized with luminescent $sp^3$ defects. Mobile excitons can decay radiatively ($E_{11}$ emission) or non-radiatively by quenching at nanotube ends and defect sites (Q). Localized $sp^3$ defects result in red-shifted emission ($E_{11}^*$ or $E_{11}^{*-}$). Normalized (to $E_{11}$) PL spectra of polymer-sorted (6,5) SWCNTs functionalized **(b)** with 4-nitrobenzenediazonium tetrafluoroborate ($E_{11}^*$ defects), **(c)** with 2-iodoaniline ($E_{11}^{*-}$ defects), and **(d)** with both $E_{11}^*$ and $E_{11}^{*-}$ defects using a sequential reaction scheme. **(e)** Schematic of absolute PLQY measurements with an integrating sphere.

The increasing defect emission intensities in relation to the intrinsic $E_{11}$ emission (~1000 nm) reflect the rising number of defects, but this ratio strongly depends on excitation power[14, 18] and cannot provide an absolute number. In contrast to that, absolute PLQY values of functionalized SWCNT dispersions can be used to calculate the density of luminescent $sp^3$ defects within the framework of the diffusion-limited contact quenching (DLCQ) model. The DLCQ model assumes that excitonic $E_{11}$ emission is governed by exciton diffusion and non-radiative decay at stationary quenching sites.[13] Within this model, luminescent $sp^3$ defects represent an additional relaxation pathway competing for mobile excitons and resulting in a lower $E_{11}$ PLQY of the functionalized SWCNTs ($\eta^*$) compared to pristine SWCNTs ($\eta$) as introduced by Miyauchi *et al.*[10] The ratio ($\eta/\eta^*$) can be used to calculate the density of luminescent defects $n_d$ [μm$^{-1}$] according to

$$n_d = \sqrt{\frac{\pi}{2\,\eta\,D\,\tau_{rad}}} \left( \sqrt{\frac{\eta}{\eta^*}} - 1 \right) \qquad (1)$$

where $D$ is the exciton diffusion constant and $\tau_{rad}$ is the radiative lifetime of the $E_{11}$ exciton. The values for $D$ and $\tau_{rad}$ were taken from previous experimental studies on (6,5) SWCNTs[13,33] (for details see the **Supporting Information**). Absolute PLQY values of SWCNT dispersions can be obtained from the direct measurement of absorbed and emitted photons in an integrating sphere in comparison to a reference sample (cuvette with solvent, see **Figure 1e** and **Supporting Information** for experimental details) as described previously.[34] The spectral contributions of the intrinsic excitonic $E_{11}$ emission and defect emission ($E_{11}^*$, $E_{11}^{*-}$) are separated and the defect density is calculated using eq (1) and the $E_{11}$ PLQY. Based on the uncertainties of the PLQY measurements and the error margins of the reported $D$ and $\tau_{rad}$ values, a relative uncertainty of the defect density of about 15% can be estimated. **Figure 2a** shows the $E_{11}$ and $E_{11}^*$ contributions to the total PLQY versus the calculated defect densities for (6,5) SWCNTs functionalized with different concentrations of DzNO$_2$. Pristine (6,5) SWCNTs exhibit a total PLQY of ~2% in dispersion. Low levels of luminescent defects (up to

~10 µm$^{-1}$) increase the total PLQY by a factor of 2 followed by a strong reduction of the total PL yield at higher degrees of functionalization, in agreement with previous studies.[14, 18]

A major drawback of this approach is the necessity of an experimental setup with an integrating sphere for determining the PLQY in combination with precisely defined measurement conditions to prevent distortions due to photon reabsorption.[35] In contrast, determining the relative integrated D mode ($I_D$, 1200 - 1400 cm$^{-1}$) to G$^+$ mode intensity ($I_{G^+}$, 1560 - 1640 cm$^{-1}$) from Raman spectra of the corresponding drop-cast nanotube films is straightforward and very reliable. As shown in **Figure 2b**, the D mode intensity increases with the degree of functionalization, being indicative of the number of $sp^3$ carbon atoms. **Figure 2c** confirms a linear correlation of the integrated Raman signal ratio ($I_D/I_{G^+}$) with the calculated defect densities obtained from the E$_{11}$ PLQY data in **Figure 2a**.

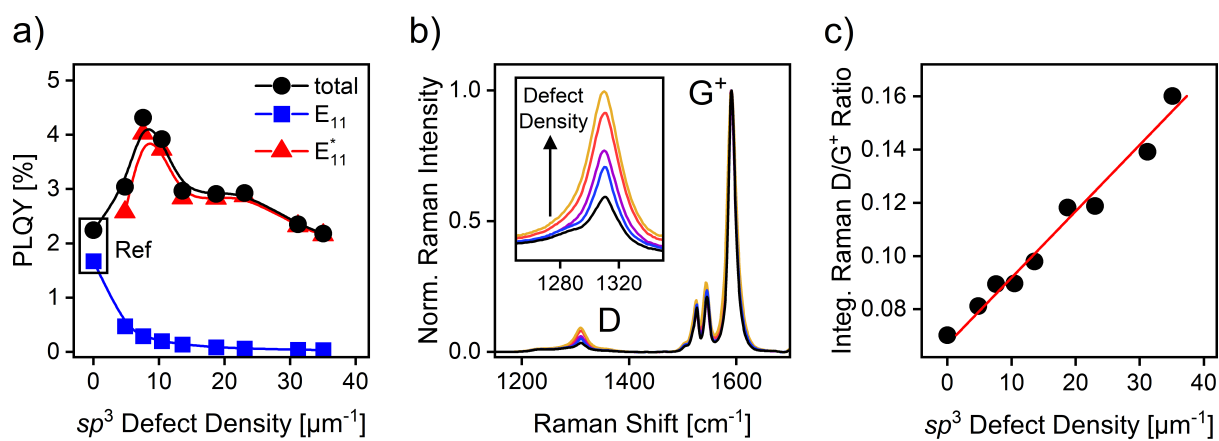

**Figure 2. (a)** PLQY data with spectral contributions of the E$_{11}$ (without sidebands) and E$_{11}$* emission at different defect densities for (6,5) SWCNTs functionalized with DzNO$_2$. Lines are guides to the eye. **(b)** Normalized Raman spectra of $sp^3$-functionalized (6,5) SWCNTs with a zoom-in on the D mode region as an inset. **(c)** Correlation of the integrated Raman D/G$^+$ ratio and defect density calculated from E$_{11}$ PLQY with linear fit (red line, R² = 0.98).

The linear correlation in **Figure 2c** should enable a direct evaluation of the number of $sp^3$ defects based only on Raman spectra of functionalized (6,5) SWCNTs. However, this metric does not take into account the variations in initial quality of nanotubes before functionalization and thus the variability of the Raman D/$G^+$ ratios of the pristine samples. As $sp^3$ functionalization adds luminescent defects to a SWCNT lattice that already contains a certain number of defects depending on starting material and processing, the absolute Raman D/$G^+$ ratios cannot be used to determine the number of introduced $sp^3$ defects. Hence, we propose to use the difference between the integrated Raman D/$G^+$ ratios of the pristine and functionalized sample, *i.e.*, $\Delta(I_D/I_{G^+})$, as a suitable metric for the quantification of defects introduced by functionalization. Note that we use the integrated Raman D/$G^+$ ratio instead of just the peak intensity ratio because it provides more reliable and reproducible values, especially for small changes in defect density.

This differential integrated Raman D/$G^+$ ratio enables comparison between different batches of nanotubes and different functionalization methods. **Figure 3** shows a summary of different batches of functionalized (6,5) SWCNTs (see **Supporting Information**, **Figures S1-S3** for detailed PLQY data, PL and Raman spectra). The linear correlation of $\Delta(I_D/I_{G^+})$ with the defect density extracted from the PLQY data holds for different defect densities, various SWCNT batches, $E_{11}$* and $E_{11}$*$^-$ defects as well as sequential functionalization to create both defects.

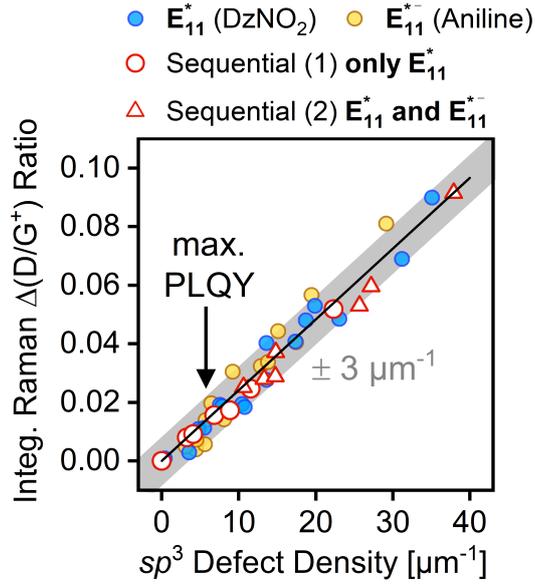

**Figure 3.** Differential integrated Raman D/G$^+$ ratio versus calculated defect densities for different batches of polymer-wrapped (6,5) SWCNTs with $E_{11}^*$, $E_{11}^{*-}$, and both defect configurations, including linear fit (black solid line, $R^2 = 0.98$) and estimated error margin (± 3 defects μm$^{-1}$ shaded in gray). Arrow: defect densities for maximum PLQY (4 – 8 μm$^{-1}$).

A linear fit to the compiled data yields the following simple expression for the density of introduced $sp^3$ defects in (6,5) SWCNTs:

$$n_d = (414 \pm 11)\ \mu m^{-1} \cdot \Delta\left(\frac{I_D}{I_{G^+}}\right) \quad (2)$$

This equation is valid across a wide range of relevant defect densities (2 – 40 defects μm$^{-1}$) and, in particular, covers the defect densities associated with maximum total PLQYs (i.e., 4 – 8 defects μm$^{-1}$).

To exclude any potential influence of the laser excitation power on the integrated Raman D/G$^+$ ratios and thus extracted defect densities, Raman spectra of functionalized (6,5) SWCNTs were recorded under identical conditions but at different excitation power densities (see **Figure S4, Supporting Information**). No significant changes of the integrated Raman D/G$^+$ ratios were observed, and values obtained at typical laser power densities for Raman spectroscopy of (6,5) SWCNTs should be comparable between different Raman spectrometers as well. Thus, this simple metric enables a quick and precise characterization of functionalized SWCNTs, and could be used for a reliable comparison of experiments with functionalized nanotubes in different laboratories using different experimental setups. It could also be applied to quickly identify a batch of functionalized nanotubes that is most likely to produce the strongest NIR emission upon excitation.

Note that other possible metrics were also tested, such as the integrated defect-to-$E_{11}$ absorption and emission ratios. In general, absorption ratios are rather unreliable due to the very low absorbance values for $E_{11}*$ and $E_{11}*^-$ transitions at low defect densities even for fairly concentrated dispersions (see **Figure S5**, **Supporting Information**). No clear correlation with the calculated defect densities could be identified across different SWCNT batches and defect configurations (see **Figure S6**, **Supporting Information**). The more commonly employed defect-to-$E_{11}$ emission ratio, which can assess the relative degree of $sp^3$ functionalization of SWCNTs,[20, 36-37] is not applicable across different batches and functionalization methods due to the non-linear and variable dependence of $E_{11}$ and defect emission on excitation power (see **Figure S7**, **Supporting Information**).[18, 38]

The demonstrated cross-correlation of integrated Raman $D/G^+$ ratios and defect densities calculated from $E_{11}$ PLQY data enables a simple evaluation of $sp^3$ defect densities. However, even the DLCQ model only quantifies the number of defects indirectly and relies heavily on correct values for the exciton diffusion constant and radiative lifetime. In contrast, PL spectra of individual functionalized SWCNTs at cryogenic temperature (cryo-PL) allow for each luminescent defect to be counted as a separate emission peak,[22, 39] assuming that each distinguishable peak within the $E_{11}*$ or $E_{11}*^-$ spectral emission range corresponds to precisely one $sp^3$ defect of the respective binding configuration. Different defect emission intensities only reflect the integrated probability for exciton relaxation in a given defect state, as previously reported for the intrinsic $E_{11}$ transition.[22]

To cross-check the calculated defect densities from PLQY measurements, two samples of (6,5) SWCNTs that were functionalized with low and medium densities of $E_{11}$* defects (spectral region 1100 – 1220 nm) were produced and PL spectra at 4.6 K from a large number of individual nanotubes embedded in a polystyrene matrix were statistically analyzed (see **Figure 4**). At low calculated defect densities of ~4 µm$^{-1}$, only few defect emission peaks were found in over 50 representative spectra (see **Figures 4a** and **S8**, **Supporting Information**). At medium defect densities (~8 µm$^{-1}$) significantly more defect PL peaks were identified on more than 40 single SWCNTs (see **Figures 4b** and **S9**, **Supporting Information**). Some $E_{11}$*$^-$ defects (spectral region 1220 – 1360 nm) were found for medium defect densities, which is consistent with literature reports for (6,5) SWCNTs functionalized with $DzNO_2$.[16-17, 40] As both defect configurations contribute equally to $E_{11}$ quenching within the DLCQ model, and the developed quantification metric does not depend on the binding configuration, all defect peaks were included in the statistical analysis. The respective histograms for the number of defects per nanotube at low and medium $sp^3$ defect densities are shown in **Figures 4c** and **4d**. The defect densities calculated from PLQY data and the average defect densities obtained from histograms are in good agreement. However, while calculated defect densities are given per µm of SWCNT, cryo-PL spectra show defects on individual nanotubes with unknown length. Hence, the length distribution of $sp^3$-functionalized SWCNTs needs to be considered for a thorough comparison. For this purpose, atomic force micrographs of nanotubes from the same dispersion of functionalized SWCNTs as those used in cryo-PL spectroscopy were recorded and statistically analyzed (see **Figures S8** and **S9**, **Supporting Information**). Both pristine and functionalized (6,5) SWCNTs exhibited average lengths of ~1.6 µm. This length distribution suggests that the number of defects per micrometer obtained via cryo-PL measurements is actually slightly lower than that extracted from PLQY measurements (possibly due to a selection bias toward less bright spots to avoid bundles and not all introduced defects being bright, see **Supporting Information**) but still within the margin of error ($\pm 3$ μm$^{-1}$) established in eq (2).

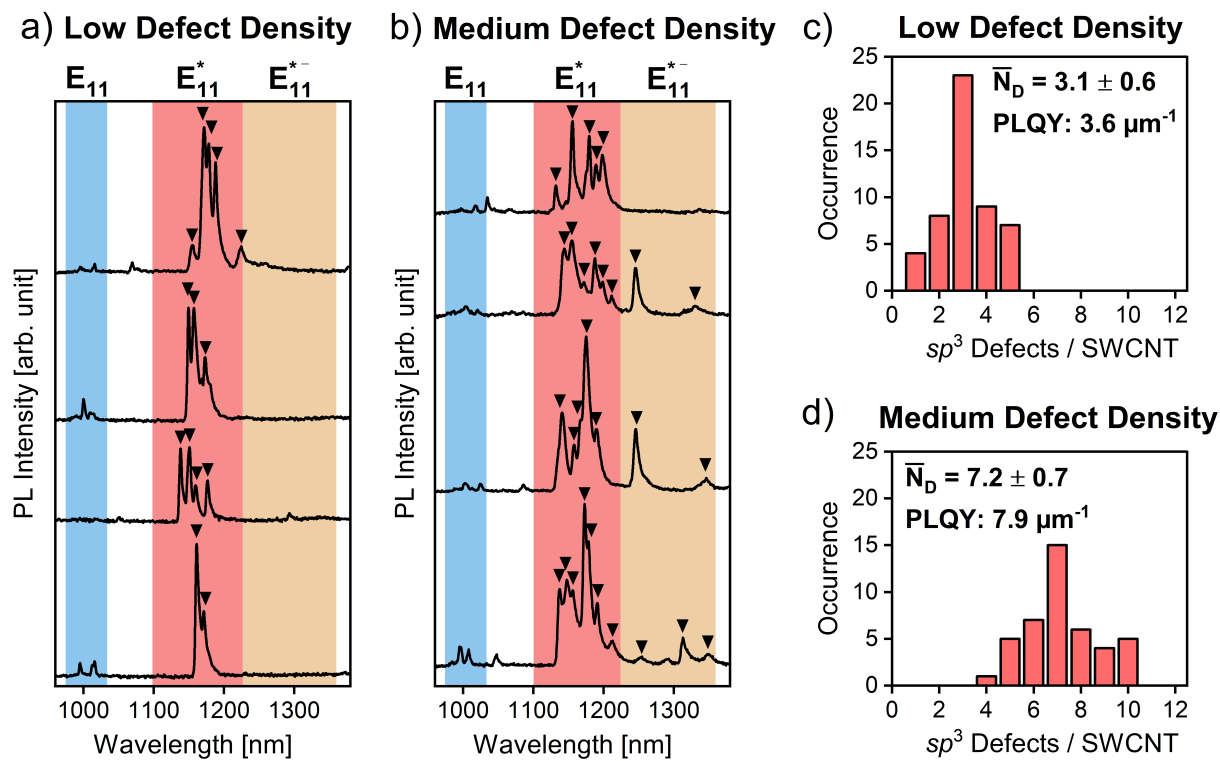

**Figure 4.** Low-temperature (4.6 K) single-SWCNT PL spectra of (6,5) SWCNTs in a polystyrene matrix functionalized with **(a)** low and **(b)** medium defect densities. Black triangles indicate individual defect emission peaks, spectral regions are highlighted for $E_{11}$ (blue), $E_{11}^*$ (red), and $E_{11}^{*-}$ (orange) emission. Defect peak histograms of **(c)** 51 functionalized (6,5) SWCNTs with defect density 3.6 μm$^{-1}$ and **(d)** 43 functionalized (6,5) SWCNTs with defect density 7.9 μm$^{-1}$ as calculated from PLQY data.

Although eq (2) was derived from measurements of long, polymer-wrapped (6,5) SWCNTs functionalized in organic solvents, the method is also applicable to short nanotubes and aqueous dispersions of SWCNTs. We used DzNO$_2$ to functionalize polymer-sorted (6,5) SWCNTs that were intentionally shortened to average lengths of 0.46 µm by tip sonication (**Supporting Information**, **Figures S10** and **S11**) and (6,5) SWCNTs sorted by aqueous two-phase extraction (ATPE), stabilized by surfactants in water (see **Methods** and **Figure S12**, **Supporting Information**). For both, a linear correlation between the differential Raman D/G$^+$ ratios and calculated defect densities in agreement with eq (2) was found (see **Figure 5a**). Moreover, nearly monochiral dispersions of polymer-sorted (7,5) SWCNTs were functionalized with DzNO$_2$ (see **Methods** and **Figure S13**, **Supporting Information**) in a first attempt to expand our approach to other nanotube species. Due to the lower reactivity of (7,5) SWCNTs compared to (6,5) SWCNTs,[18] a maximum absolute defect density of ~20 µm$^{-1}$ was achieved. Nevertheless, the presented quantification method was also applicable to (7,5) SWCNTs (wrapped with polydioctylfluorene, PFO) as shown in **Figure 5b**, although with some deviations in the precise slope.

This is particularly interesting as resonant Raman measurements of (7,5) SWCNTs are performed at a different excitation wavelength (633 nm) compared to (6,5) SWCNTs (532 nm). For point-like defects in graphene, Cançado *et al.* reported a strong dependence of the Raman D/G intensity ratio on excitation laser wavelength, which was included in the expression for the average defect distance. However, for large defect distances the influence of the excitation wavelength on the absolute Raman D/G intensity ratio becomes negligible.[23] The mean defect distances in *sp*$^3$-functionalized SWCNTs relevant for most applications and investigated here (20-300 nm, see **Figure S14**, **Supporting Information**) are much larger than those considered in studies of defective graphene (5-30 nm).[25] Nevertheless, it remains to be tested what impact the Raman laser excitation wavelength has on SWCNTs with larger diameters.

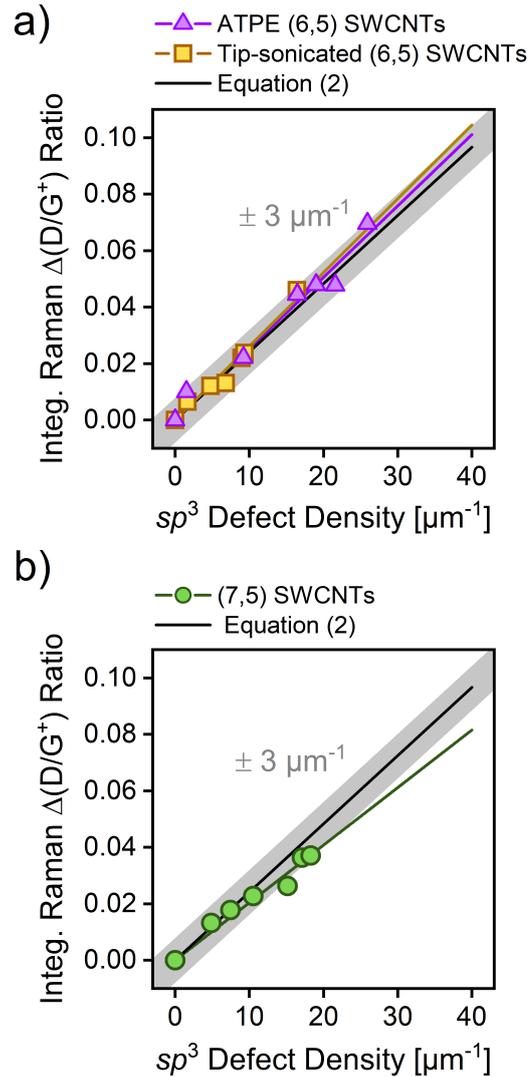

**Figure 5.** Differential integrated Raman D/G$^+$ ratio versus calculated defect densities and linear fits for **(a)** aqueous dispersions of ATPE-sorted (6,5) SWCNTs and tip-sonicated, polymer-wrapped (6,5) SWCNTs and for **(b)** polymer-wrapped (7,5) SWCNTs. All functionalization steps were performed with DzNO$_2$.

One direct application of the presented quantification method is the experimental determination of the oscillator strengths of E$_{11}$* and E$_{11}$*$^−$ defects. Integration of the defect peak areas in the NIR absorption spectra of functionalized SWCNTs with different defect densities yielded the integrated molar absorptivities and absorption cross sections of the E$_{11}$* and E$_{11}$*$^−$ states (see

**Table 1** and **Figure S15**, **Supporting Information**). The oscillator strength of an optical transition was calculated using

$$f = \frac{4\,\varepsilon_0\,c^2\,m_e\ln(10)}{N_A\,e^2} \int \varepsilon_D\,d\tilde{v} \qquad (3)$$

where $\varepsilon_0$ denotes the vacuum permittivity, $c$ is the speed of light, $m_e$ is the electron mass, $N_A$ is Avogadro's number, $e$ is the elementary charge, $\varepsilon_D$ is the molar extinction coefficient of the $E_{11}$* or $E_{11}$*⁻ transition, and $\tilde{v}$ is the wavenumber.[41]

For $E_{11}$* and $E_{11}$*⁻ defects, oscillator strengths of (3.5 ± 0.6) and (1.1 ± 0.3) per defect were obtained, respectively. The error margins for the oscillator strengths of $E_{11}$* and $E_{11}$*⁻ are mainly due to the low total absorbances of the *sp*³ defects and the corresponding uncertainties of the integrated defect absorption (see **Figure S5**, **Supporting Information**). While density functional theory calculations predicted larger oscillator strengths (by a factor of 3 – 5),[16-17] they also suggested a reduction of the oscillator strength with a greater red-shift of the defect emission, which is in agreement with our findings. It is important to note that the discussed spectroscopic metrics provided here are given per defect site and are not directly comparable to literature data on the $E_{11}$ oscillator strengths of (6,5) SWCNTs (~0.01 per carbon atom).[42] As electron-hole correlation lengths are still on the order of 1 nm despite localization at defect sites,[43] ~100 carbon atoms are presumed to contribute to the *sp*³ defect oscillator strength. Although the acquired data is not accurate enough to corroborate an increase of oscillator strength by a factor of two upon exciton trapping as proposed by Miyauchi *et al.*,[10] our findings suggest that the oscillator strengths of *sp*³ defects are at least on the same order of magnitude as for the $E_{11}$ transition.

**Table 1.** Integrated absorptivity, integrated absorption cross section, and oscillator strength per defect for E$_{11}$* and E$_{11}$*$^-$ defect configurations, obtained from NIR absorption data of *sp*$^3$-functionalized (6,5) SWCNTs.

| Defect configuration | Integrated absorptivity $\int \varepsilon_D \, d\lambda$ [cm$^{-1}$ nm L mol$^{-1}$] | Integrated absorption cross section $\int \sigma \, d\tilde{v}$ [cm] | Oscillator strength $f$ |
|---|---|---|---|
| $\mathbf{E^*_{11}}$ | $(7.8 \pm 1.4) \cdot 10^7$ | $(3.1 \pm 0.5) \cdot 10^{-12}$ | $3.5 \pm 0.6$ |
| $\mathbf{E^{*-}_{11}}$ | $(2.5 \pm 0.4) \cdot 10^7$ | $(1.0 \pm 0.3) \cdot 10^{-12}$ | $1.1 \pm 0.3$ |

In conclusion, we have developed a straightforward and reliable method to quantify the number of *sp*$^3$ defects in (6,5) SWCNTs using resonant Raman spectroscopy. By establishing a linear correlation between the differential integrated Raman D/G$^+$ ratio and *sp*$^3$ defect densities calculated from PLQYs, the number of added defects per µm nanotube can be obtained within an error margin of ± 3 defects µm$^{-1}$. This method is suitable for SWCNTs functionalized with E$_{11}$*, E$_{11}$*$^-$, or both defect configurations, independent of the type of dispersion (polymer-wrapped in organic solvent or surfactant-stabilized in water) and nanotube length of the starting material. A statistical analysis based on PL spectra of individual (6,5) SWCNTs at cryogenic temperature provided direct access to the number of defects per nanotube, which roughly matched the densities calculated from Raman spectra. From these data the oscillator strengths of the E$_{11}$* and E$_{11}$*$^-$ defects were determined experimentally, confirming the predicted decrease in oscillator strength with optical trap depth of the defects. The applicability of the presented quantification method also extends to other nanotube species as demonstrated for (7,5) SWCNTs. However, additional resonant Raman and PLQY data of functionalized SWCNTs with different diameters will be required to obtain a universal expression similar to that for defects in graphene.

# Acknowledgments

This project has received funding from the European Research Council (ERC) under the European Union's Horizon 2020 research and innovation programme (Grant Agreement No. 817494 "TRIFECTs"). B. S. F. and H. L. acknowledge funding from the Deutsche Forschungsgemeinschaft (DFG) under research grants FL 834/2-2, FL 834/5-1, FL 834/7-1, FL 834/9-1 and FL 834/13-1.

# Supporting Information

# Supplementary Methods

## Preparation of SWCNT Dispersions

### Selective Dispersion of (6,5) and (7,5) SWCNTs

Dispersions of (6,5) SWCNTs were prepared by shear-force mixing (SFM, Silverson L2/Air, 10230 rpm, 20 °C, 72 h) of CoMoCAT raw material (Sigma-Aldrich, Charge No. MKCJ7287) in a solution of poly[(9,9-dioctylfluorenyl-2,7-diyl)-*alt*-(6,6′-(2,2′-bipyridine))] (PFO-BPy, American Dye Source, $M_W$ = 40 kg·mol$^{-1}$, 0.5 g·L$^{-1}$) in toluene according to a protocol by Graf *et al.*[1] Dispersions of (7,5) SWCNTs were selectively dispersed by shear-force mixing in a solution of poly[(9,9-dioctylfluorenyl-2,7-diyl)] (PFO, Sigma-Aldrich, $M_W$ > 20 kg·mol$^{-1}$, 0.9 g·L$^{-1}$) in toluene.

After removal of unexfoliated material by centrifugation (Beckman Coulter Avanti J26SXP centrifuge) at 60000 × *g* for 1 h, the collected supernatant was passed through a poly(tetra-fluoroethylene) (PFTE) syringe filter (pore size 5 µm). Excess polymer was removed by vacuum filtration through PTFE membrane filters (pore size 0.1 µm) and washing of the filter cakes with toluene at 80 °C, followed by redispersion in fresh toluene using bath sonication.

### Tip Sonication of (6,5) SWCNTs

Toluene dispersions of SFM (6,5) SWCNTs were tip-sonicated (Sonics, Vibracell VXC-500) with a tapered microtip at 35 % amplitude with 8 seconds on and 2 seconds off pulses at 5 °C for 48 h. The resulting dispersions were centrifuged at 60000 × *g* for 45 min, and the collected supernatant was used for characterization and further processing.

## Characterization Methods

### Absorption Spectroscopy

Baseline-corrected absorption spectra were recorded using a Cary 6000i UV-Vis-NIR absorption spectrometer (Varian, Inc.) and cuvettes with 1 cm path length. A scattering background $S(\lambda) = S_0 e^{-b\lambda}$ was fitted and subtracted from the acquired absorption spectra.[2-3]

### Atomic Force Micrographs

Atomic force micrographs were recorded using a Bruker Dimension Icon atomic force microscope (AFM) in ScanAsyst$^{TM}$ mode under ambient conditions. SWCNTs were spin-coated (2000 rpm, 60 s) from toluene dispersions with an optical density (OD) of 0.15 cm$^{-1}$ at the $E_{11}$ transition onto cleaned native silicon wafers. The wrapping polymer was removed by rinsing with tetrahydrofuran (THF) and isopropanol. Length distributions were analyzed using Gwyddion 2.6.

**Raman Spectroscopy of SWCNTs**

Raman spectra of pristine and $sp^3$-functionalized SWCNTs were collected with a Renishaw inVia Reflex confocal Raman microscope in backscattering configuration equipped with a 50× long working distance objective (N.A. 0.5). Dispersions of SWCNTs were drop-cast on glass substrates (Schott AF32eco) and rinsed with THF and isopropanol (toluene-based dispersions) or ultrapure water (aqueous dispersions). Near-resonant excitation of the samples was performed with 532 nm and 633 nm lasers for (6,5) and (7,5) SWCNTs, respectively. For each sample >3600 spectra were averaged and baseline-corrected.

**Photoluminescence Spectroscopy**

Room-temperature photoluminescence (PL) spectra of SWCNT dispersions were acquired using the wavelength-filtered output of a picosecond-pulsed supercontinuum laser (Fianium WhiteLase SC400, repetition rate 20 MHz, pulse width ~6 ps) focused on the samples using a 50× NIR-optimized objective (N.A. 0.65, Olympus). Scattered excitation light was blocked using appropriate long-pass filters. The PL emission of SWCNT dispersions was collected with an Acton SpectraPro SP2358 spectrometer (grating blaze, 1200 nm, 150 lines·mm$^{-1}$) and a liquid nitrogen-cooled InGaAs line camera (Princeton Instruments OMA V:1024). Resonant excitation at the $E_{22}$ transition was performed at wavelengths of 575 nm and 652 nm for (6,5) and (7,5) SWCNT dispersions, respectively.

**PL Quantum Yield Measurements**

The PL quantum yields (PLQYs) of pristine and $sp^3$-functionalized SWCNTs in dispersion were determined using an absolute method as reported previously.[1] SWCNT dispersions were diluted to an OD of 0.2 cm$^{-1}$ at the $E_{11}$ transition to minimize re-absorption. For all measurements, 1 mL of the analyte was filled into a cuvette (Hellma Analytics, QX type), which was placed in an integrating sphere (LabSphere, Spectralon coating). SWCNTs were excited resonantly at the $E_{22}$ transition by the wavelength-filtered output of a picosecond-pulsed supercontinuum laser source (Fianium WhiteLase SC400). The light exiting the integrating sphere was coupled into an Acton SpectraPro SP2358 spectrometer using an optical fiber. Spectra were acquired using a liquid nitrogen-cooled InGaAs line camera (Princeton Instruments OMA V:1024).

PLQYs were calculated as the number ratio of emitted ($N_{em}$) to absorbed photons ($N_{abs}$) according to:

$$PLQY = \frac{N_{em}}{N_{abs}} \qquad (1)$$

For SWCNT dispersions, a value proportional to $N_{abs}$ is obtained by integration of the laser signal intensity relative to that of a solvent reference sample (note, the product of intensity and wavelength is used for the integral to be proportional to the number of photons). Equally, integration of recorded PL spectra relative to the PL signal of the solvent provides a value proportional to $N_{em}$, allowing for the calculation of the PLQY.[4] Detector efficiency and absorption characteristics of optical components were accounted for by acquisition of calibration spectra employing a broadband light source (Thorlabs SLS201/M). NIR absorption of the solvent was corrected for by measuring the lamp spectrum while a cuvette filled with the respective solvent was placed inside the integrating sphere. Correction measurements for absorption spectra in the spectral region of the $E_{22}$ absorption were performed without the cuvette in the integrating sphere. Prior to integration, every spectrum was divided by the correction function calculated from the ratio of the recorded and theoretical lamp spectrum. The error of the PLQY determination primarily results from uncertainties for the laser absorption measurements and is estimated to be 10%.

**Low-Temperature Single-SWCNT PL Spectroscopy**

Low-temperature PL spectroscopy of individual SWCNTs was performed at 4.6 K using a closed-cycle liquid helium-cooled optical cryostat (Montana Instruments, Cryostation s50). Dispersions of SWCNTs were diluted with a solution of polystyrene (Polymer Source Inc., $M_W$ = 230 kg·mol$^{-1}$) in toluene (20 g·L$^{-1}$) to an OD of 0.005 cm$^{-1}$ at the $E_{11}$ absorption peak. 30 µL of the obtained mixture were spin-coated (2000 rpm, 60 s) onto a glass substrate coated with 150 nm of gold.

The output of a continuous wave laser diode (Coherent, Inc., OBIS 640 nm, 1 mW) was focused onto the sample using a NIR-optimized 50× long-working distance objective (Mitutoyo, N.A. 0.42). Scattered laser light was blocked by appropriate long-pass filters. Low-temperature PL spectra were recorded with a thermoelectrically cooled two-dimensional InGaAs camera array (Princeton Instruments, NIRvana 640ST) coupled to a grating spectrograph (Princeton Instruments, IsoPlane SCT-320) using a grating with 85 grooves·mm$^{-1}$ and 1200 nm blaze. Very bright or asymmetrical emission spots were disregarded for PL measurements as they likely originate from SWCNT bundles instead of individual nanotubes. Only those peaks were included in the statistical analysis that could be unambiguously identified as single-defect emission based on the sharp, asymmetric peak shape typically observed for polymer-wrapped

SWCNTs at cryogenic temperatures. For any analyzed spectrum the precise number of defects might be underestimated by 1-2 at most.

## sp³ Functionalization Protocol for SWCNTs

**Functionalization of Polymer-Wrapped SWCNTs with 4-Nitrobenzenediazonium Tetrafluoroborate**

Dispersions of (6,5) SWCNTs wrapped with PFO-BPy were functionalized with 4-nitrobenzenediazonium tetrafluoroborate ($DzNO_2$, Sigma-Aldrich) according to a protocol by Berger *et al.*[5] The reaction was carried out in a mixture of toluene and acetonitrile, using 18-crown-6 (Sigma-Aldrich) as phase-transfer agent. Variations of the defect density were achieved by adjustment of the diazonium salt concentration. As the first step of the functionalization procedure, a solution of 18-crown-6 in toluene was prepared so that its final concentration in the reaction mixture after addition of all remaining reactants amounted to 7.6 mmol·L$^{-1}$. Subsequently, an appropriate volume of (6,5) SWCNT dispersion in toluene was added to achieve a final SWCNT concentration of 0.36 mg·L$^{-1}$. This concentration is equivalent to an OD of 0.2 cm$^{-1}$ at the $E_{11}$ transition and was chosen to reduce the formation of SWCNT bundles. Acetonitrile was then added to the reaction mixture so that a ratio of toluene to acetonitrile of 80:20 vol-% would be achieved in the final reaction mixture. A stock solution of $DzNO_2$ in acetonitrile (5 g·L$^{-1}$) was prepared in an amber vial. Depending on the desired concentration in the final reaction mixture, an appropriate amount of the stock solution was added. In this work, a minimum concentration of 20 g·L$^{-1}$ and a maximum concentration of 1000 mg·L$^{-1}$ were chosen. After storage in the dark for 16 h, the reaction mixture was filtered through a PTFE membrane (Merck Millipore JVWP, pore size 0.1 µm). The filter cake was washed with 10 mL of acetonitrile and 5 mL of toluene to remove unreacted diazonium salt and excess wrapping polymer. The filter cake with functionalized SWCNTs was redispersed in 1 mL of pure toluene using bath sonication for 30 min.

The functionalization of (7,5) SWCNTs was performed similarly but with a concentration of 2000 mg·L$^{-1}$ of $DzNO_2$ for all samples. Defect densities were controlled by variation of the reaction time, for which a minimum of 14 h and a maximum of 208 h were chosen.

**Functionalization of Polymer-Wrapped SWCNTs with 2-Iodoaniline**

(6,5) SWCNTs were functionalized with 2-iodoaniline according to the procedure by Settele *et al.*[6] Nanotube dispersions were filtered through a PTFE membrane (Merck Millipore JVWP, pore size 0.1 μm) and washed with 10 mL of toluene to remove excess wrapping polymer. The obtained filter cakes were redispersed in 1 mL of pure toluene using bath sonication for 30 min. 2-Iodoaniline (Sigma-Aldrich) was dissolved in toluene in an amber vial equipped with a magnetic stirring bar to obtain a concentration of 29.3 mmol·L$^{-1}$ in the final reaction mixture. Subsequently, 1 mL of dimethyl sulfoxide (DMSO) and 1 mL of a solution of potassium *tert*-butoxide in anhydrous THF, prepared under nitrogen atmosphere, were added to the solution of 2-iodoaniline. The polymer-depleted dispersion of SWCNTs was added such that the OD of the reaction mixture at the $E_{11}$ transition was 0.3 cm$^{-1}$, equivalent to a (6,5) SWCNT concentration of 0.54 mg·L$^{-1}$. The final total volume of toluene in the reaction mixture amounted to 10 mL. After stirring at room temperature for time periods between 5 min and 180 min, depending on the targeted defect density, the reaction mixture was filtered through a PTFE membrane (Merck Millipore JVWP, pore size 0.1 μm). The filter cake was washed with 5 mL of methanol, 5 mL of toluene, and redispersed in 1 mL of PFO-BPy solution (0.1 g·L$^{-1}$ in toluene) by bath sonication for 30 min. The additional wrapping polymer is only added to improve colloidal stability of the dispersion for storage and characterization.

**Sequential Functionalization of Polymer-Wrapped SWCNTs**

The sequential functionalization procedure of (6,5) SWCNTs employed both protocols presented before. In the first step, SWCNTs were functionalized with DzNO$_2$ in different concentrations ranging from 20 mg·L$^{-1}$ to 1000 mg·L$^{-1}$. The obtained dispersions were then subjected to an additional functionalization step using 2-iodoaniline and a fixed reaction time of 40 min.

**ATPE Separation and Functionalization of Aqueous (6,5) SWCNT Dispersions**

The separation of (6,5) SWCNTs by aqueous two-phase extraction (ATPE) was performed according to a previously reported method.[7] Dextran (M$_W$ = 70 kDa, TCI) and poly(ethylene glycol) (PEG, M$_W$ = 6 kDa, Alfa Aesar) were used to form the two-phase system, in which CoMoCAT raw material (CHASM SG65i-L58) was separated by a diameter sorting protocol on the basis of sodium deoxycholate (DOC, BioXtra) and sodium dodecyl sulfate (SDS, Sigma-Aldrich). With a fixed DOC concentration (0.04%, w/v), the concentration of SDS was first increased to 1.1% (w/v) in order to push all species with diameters larger than (6,5) SWCNTs to the top phase for removal. Subsequently, the SDS concentration was gradually increased from 1.2% to 1.5% to collect all (6,5) SWCNT-enriched fractions. The separation of metallic and semiconducting SWCNT species was performed by addition of sodium cholate (SC, Sigma-Aldrich) and adjustment of the total surfactant concentrations to 0.9% SC, 1% SDS and <0.02% DOC, followed by addition of sodium hypochlorite (NaClO, 5 μL·mL$^{-1}$ of the 1/100th concentration, Sigma-Aldrich) as the oxidant. The sorted (6,5) SWCNTs were concentrated and adjusted to 0.1% (w/v) DOC using iterative concentration dilution cycles in a pressurized ultrafiltration stirred cell (Millipore) with a 300 kDa $M_W$ cutoff membrane. Finally, the 0.2 mL concentrated (6,5) SWCNT dispersion (in 0.1% DOC) was added to 9.8 mL of a 1% SDS solution for further processing.

For the functionalization of aqueous (6,5) SWCNT dispersions with luminescent $sp^3$ defects, a stock solution of $DzNO_2$ (0.1 mg·L$^{-1}$) in ultrapure water was prepared. Appropriate amounts of $DzNO_2$ solution and water were added to a vial with 1 mL of aqueous SWCNT dispersion, such that a final OD of 0.33 cm$^{-1}$ at the $E_{11}$ transition was obtained (corresponding to a total reaction volume of 3 mL). In this work, $DzNO_2$ concentrations between 0.005 mg·L$^{-1}$ and 0.06 mg·L$^{-1}$ were chosen. Reaction mixtures were stored overnight and sonicated for 20 min prior to characterization and further processing.

**Calculation of *sp*³ Defect Densities and Concentrations**

The number densities of luminescent $sp^3$ defects per micrometer were calculated employing the theoretical model of diffusion-limited contact quenching (DLCQ) for pristine and functionalized SWCNTs. The model is based on the competition between non-radiative quenching events and radiative relaxation of highly mobile excitons as described by Miyauchi *et al.*[8] In pristine SWCNTs, an increasing density of non-radiative quenching sites results in a drop of the intrinsic $E_{11}$ emission. Within the DLCQ model, the $E_{11}$ PLQY of pristine SWCNTs $\eta$ is expressed as

$$\eta = \frac{\pi}{2 \cdot n_q^2 \cdot D \cdot \tau_{rad}} \qquad (2)$$

where $n_q$ is the number density of quenching sites per micrometer, $D$ is the exciton diffusion constant and $\tau_{rad}$ is the radiative lifetime of the $E_{11}$ exciton. For the latter parameters, literature values of $D = (10.7\pm0.4)$ cm$^2$·s$^{-1}$ and $\tau_{rad} = (3.35 \pm0.41)$ ns were used.[9-10]

For a $sp^3$ defect density of $n_d$, the $E_{11}$ PLQY of functionalized SWCNTs $\eta^*$ can be written as:

$$\eta^* = \frac{\pi}{2 \cdot (n_q + n_d)^2 \cdot D \cdot \tau_{rad}} \qquad (3)$$

The combination of eqs (2) and (3) yields an expression for the density of luminescent $sp^3$ defects:

S-7

$$n_d = n_q \left( \sqrt{\frac{\eta}{\eta^*}} - 1 \right) = \sqrt{\frac{\pi}{2 \cdot \eta \cdot D \cdot \tau_{rad}}} \left( \sqrt{\frac{\eta}{\eta^*}} - 1 \right) \qquad (4)$$

The uncertainty of the defect density was estimated to be about 15% as a result of the error margins of the PLQY measurements and the values for $D$ and $\tau_{rad}$.

The defect concentration $c_d$ [nmol·L$^{-1}$] can be obtained from the defect density by

$$c_d = \frac{n_d \cdot A_{11}}{88 \text{ nm}^{-1} \cdot \varepsilon_{11}} \qquad (5)$$

where the geometrical factor of 88 nm$^{-1}$ is the number of carbon atoms per nanometer of (6,5) SWCNT, $A_{11}$ is the decadic absorbance at the E$_{11}$ transition, and $\varepsilon_{11}$ is the molar absorption coefficient of the E$_{11}$ transition. For the latter, a value of 6700 L·mol$^{-1}$·cm$^{-1}$ was used as determined by Streit et al.[11]

# Supplementary Figures

## (6,5) SWCNTs Functionalized with DzNO$_2$

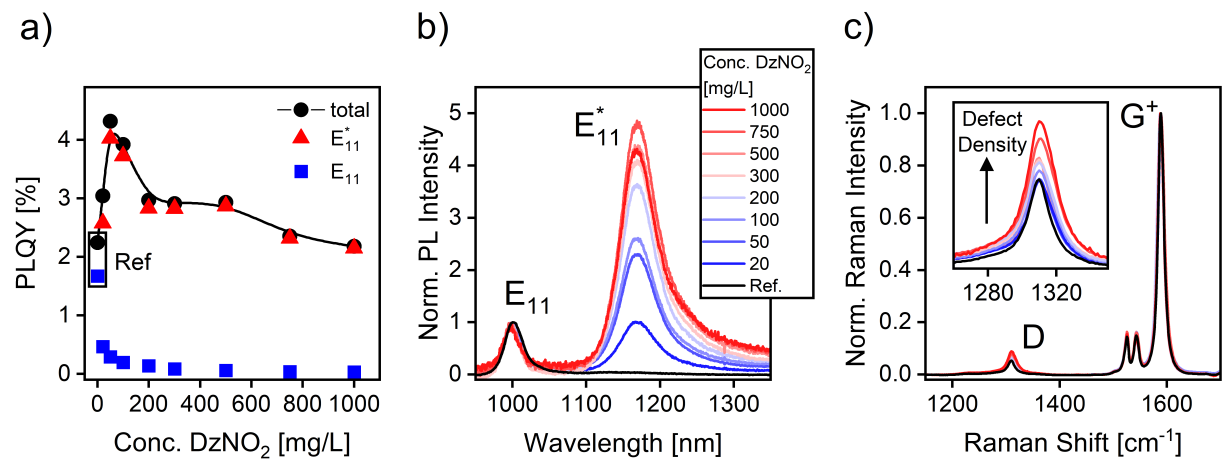

**Figure S1.** Spectroscopic data of (6,5) SWCNTs wrapped with PFO-BPy and functionalized with different concentrations of DzNO$_2$. **(a)** Spectrally separated PLQY contributions of the $E_{11}$ emission and the $E_{11}$* defect emission. The solid line is a guide to the eye. **(b)** Normalized photoluminescence spectra, acquired under pulsed excitation at the $E_{22}$ transition (575 nm, ~0.025 mJ·cm$^{-2}$). **(c)** Averaged and normalized Raman spectra and zoom-in on the D mode region, recorded with a 532 nm laser at a power density of 4.1 kW·cm$^{-2}$.

**(6,5) SWCNTs Functionalized with 2-Iodoaniline**

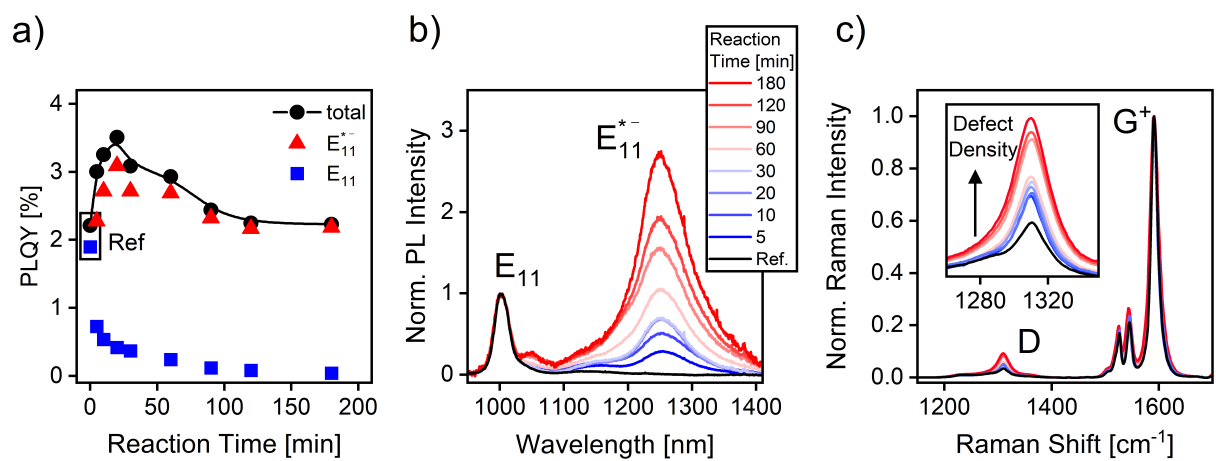

**Figure S2.** Spectroscopic data of (6,5) SWCNTs wrapped with PFO-BPy and functionalized with 2-iodoaniline with different reaction times. **(a)** Spectrally separated PLQY contributions of the $E_{11}$ emission and the $E_{11}^{*-}$ defect emission. The solid line is a guide to the eye. **(b)** Normalized photoluminescence spectra, acquired under pulsed excitation at the $E_{22}$ transition (575 nm, ~0.025 mJ·cm$^{-2}$). **(c)** Averaged and normalized Raman spectra and zoom-in on the D mode region, recorded with a 532 nm laser at a power density of 4.1 kW·cm$^{-2}$.

**Sequentially Functionalized (6,5) SWCNTs**

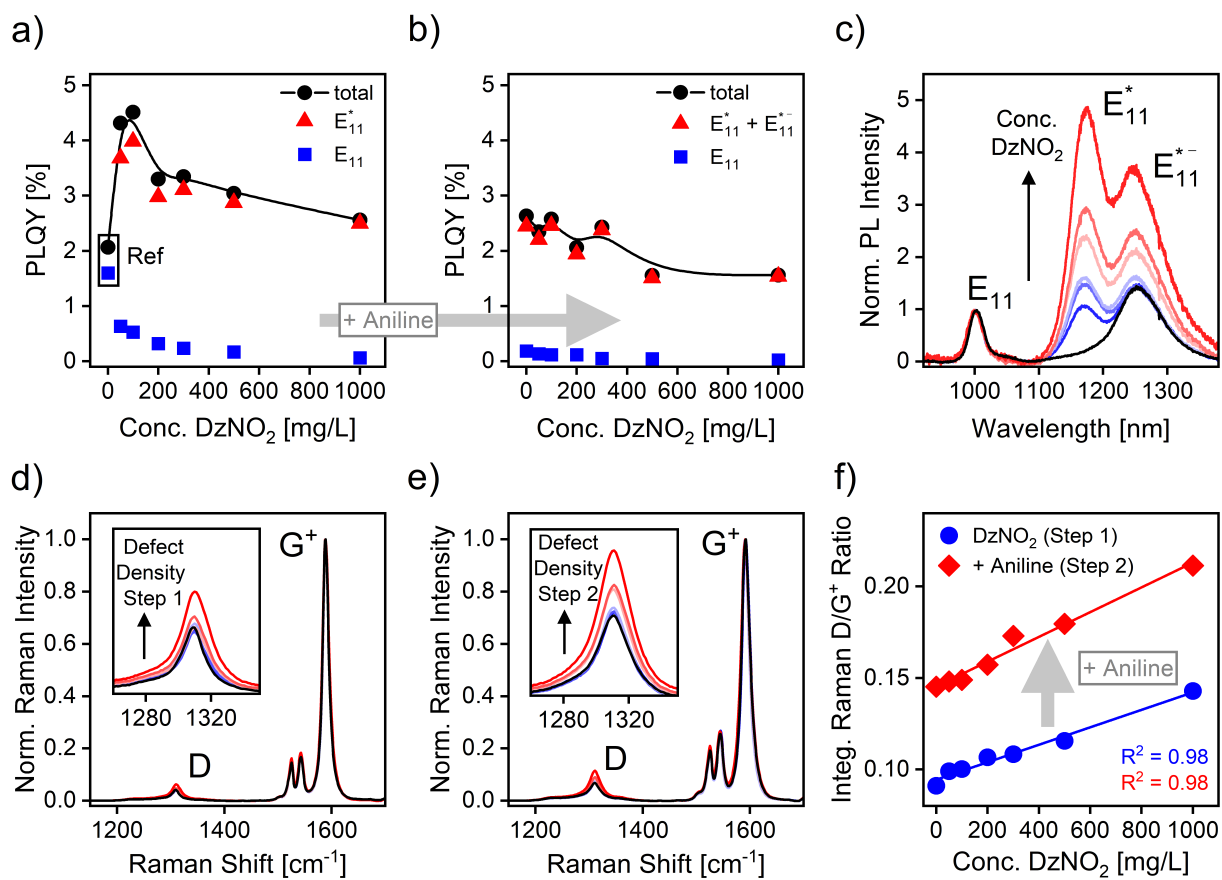

**Figure S3.** Spectroscopic data of (6,5) SWCNTs wrapped with PFO-BPy and functionalized in a sequential reaction procedure with different concentrations of DzNO$_2$ (1$^{st}$ step) and 2-iodo-aniline for 40 min (2$^{nd}$ step). **(a)** Spectrally separated PLQY contributions of the E$_{11}$ emission and the E$_{11}$* defect emission after the first functionalization step. Solid lines are guides to the eye. **(b)** Spectrally separated PLQY contributions of the E$_{11}$ emission and combined E$_{11}$*+E$_{11}$*$^-$ defect emission after the second functionalization step. **(c)** Normalized photoluminescence spectra after the second functionalization step, acquired under pulsed excitation at the E$_{22}$ transition (575 nm, ~0.025 mJ·cm$^{-2}$). Averaged and normalized Raman spectra and zoom-in on the D mode region, **(d)** after the first functionalization step and **(e)** after the second functionalization step, recorded with a 532 nm laser at a power density of 4.1 kW·cm$^{-2}$. **(f)** Correlation between Raman D/G$^+$ area ratio and diazonium salt concentration for both steps of the sequential functionalization protocol.

**Laser Excitation Power Dependence of Integrated Raman D/G$^+$ Ratio**

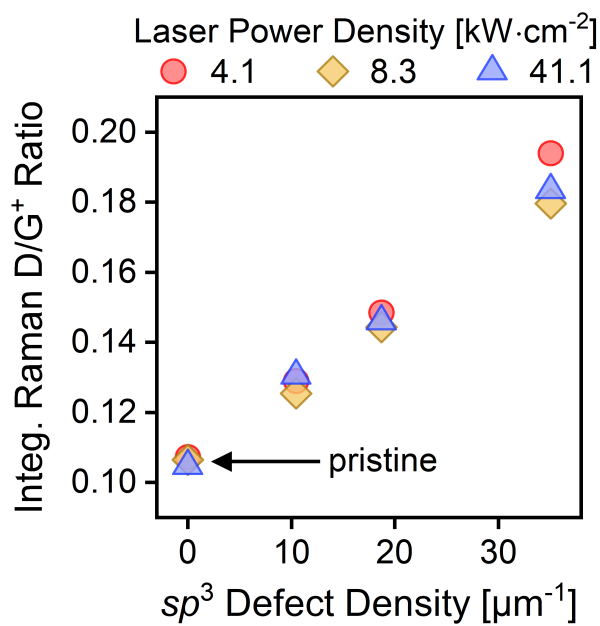

**Figure S4.** Laser excitation power density dependence of Raman D/G$^+$ ratios: Correlation between integrated Raman D/G$^+$ ratios and defect densities for (6,5) SWCNTs wrapped with PFO-BPy and functionalized with DzNO$_2$. Averaged and integrated Raman spectra were acquired with a 532 nm laser with excitation powers of 4.1 to 41.1 kW·cm$^{-2}$ with no significant changes in the integrated D/G$^+$ ratios.

**Absorption Spectra of Functionalized (6,5) SWCNTs**

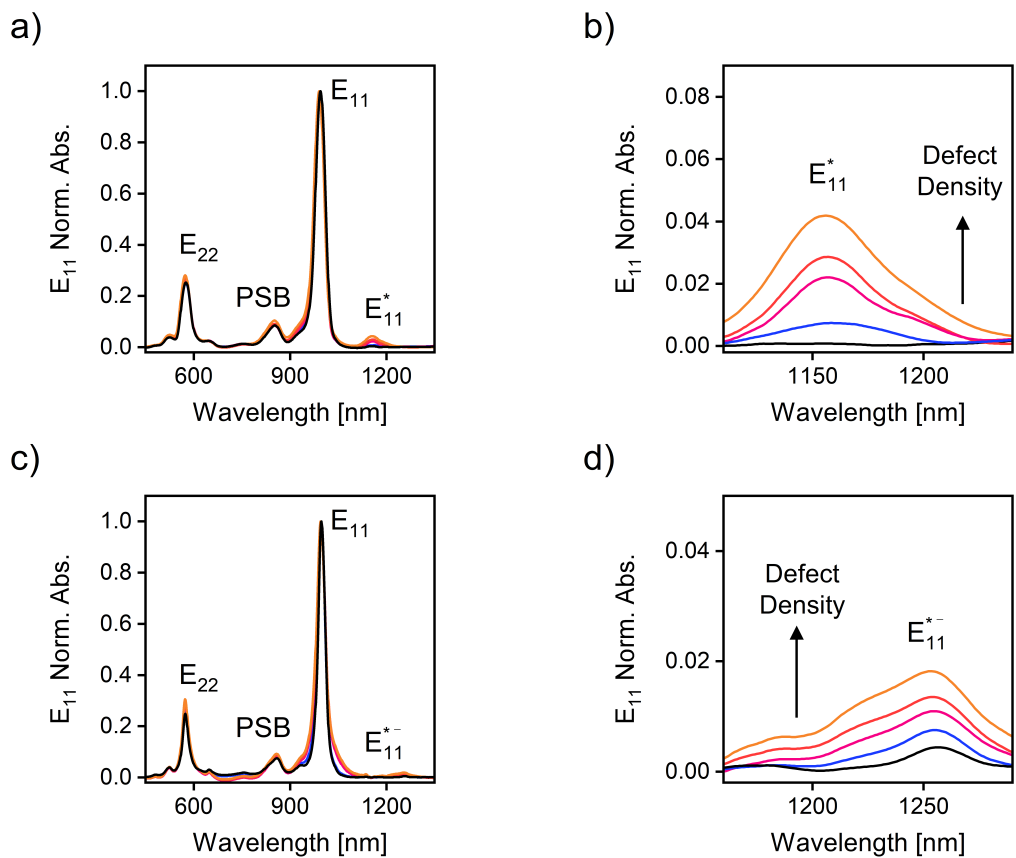

**Figure S5.** Normalized Vis-NIR absorption spectra and detail of the defect absorption feature for (6,5) SWCNTs functionalized with different densities of $E_{11}^*$ defects using $DzNO_2$ **(a, b)** and $E_{11}^{*-}$ defects using 2-iodoaniline **(c, d)** for functionalization.

**Differential Defect-to-E$_{11}$ Absorption Area Ratios for (6,5) SWCNTs**

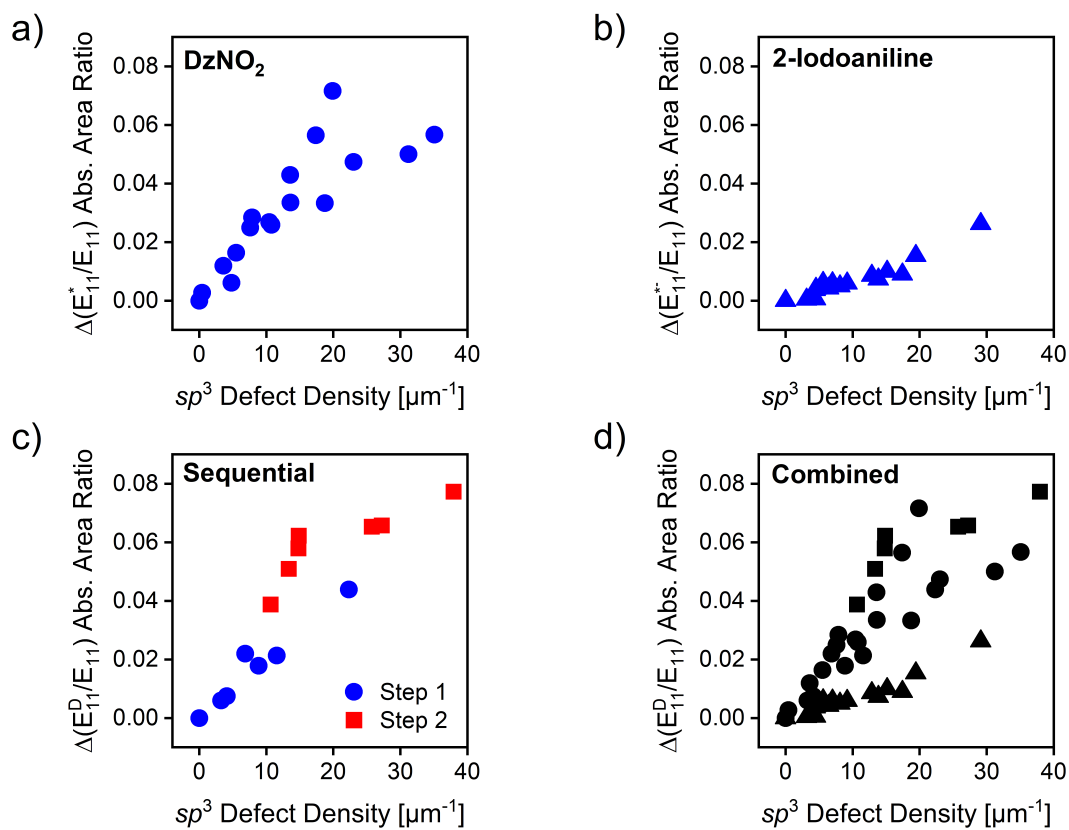

**Figure S6.** Correlation between differential defect-to-E$_{11}$ absorption area ratios and defect densities obtained from PLQYs of $sp^3$-functionalized (6,5) SWCNTs wrapped with PFO-BPy. **(a)** Data for (6,5) SWCNTs functionalized with DzNO$_2$ (E$_{11}$* defects). **(b)** Data for (6,5) SWCNTs functionalized with 2-iodoaniline (E$_{11}$*$^-$ defects). **(c)** Data for (6,5) SWCNTs sequentially functionalized with E$_{11}$* defects (1$^{st}$ step) and additional E$_{11}$*$^-$ defects (2$^{nd}$ step). E$_{11}^D$ - both E$_{11}$*$^-$ and E$_{11}$* defects. **(d)** Combined datasets showing no common correlation.

**Defect-to-E$_{11}$ PL Area Ratios for (6,5) SWCNTs**

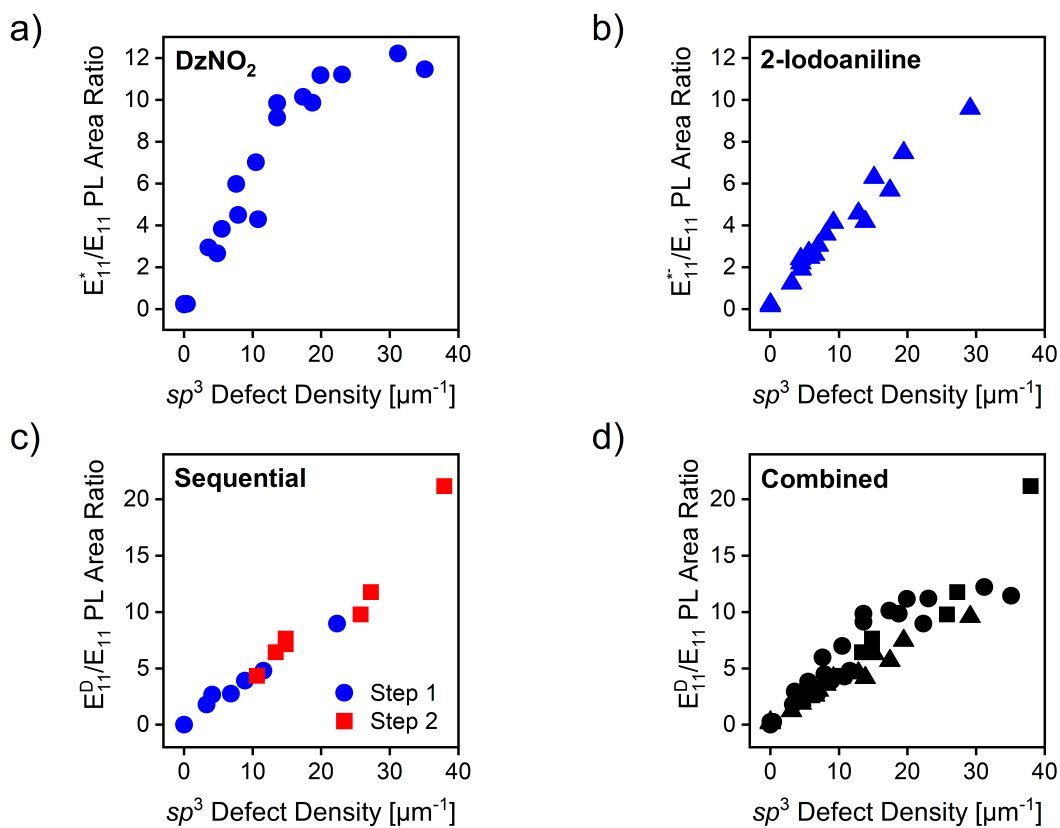

**Figure S7.** Correlation between defect-to-E$_{11}$ PL emission area ratios and defect densities obtained from PLQYs of $sp^3$-functionalized (6,5) SWCNTs wrapped with PFO-BPy. **(a)** Data for (6,5) SWCNTs functionalized with DzNO$_2$ (E$_{11}$* defects). **(b)** Data for (6,5) SWCNTs functionalized with 2-iodoaniline (E$_{11}$*$^-$ defects). **(c)** Data for (6,5) SWCNTs sequentially functionalized with E$_{11}$* defects (1$^{st}$ step) and additional E$_{11}$*$^-$ defects (2$^{nd}$ step). E$_{11}^D$ - both E$_{11}$*$^-$ and E$_{11}$* defects. **(d)** Combined datasets showing similar trends but no common linear correlation.

**Low-Temperature PL Spectra of Single Nanotubes (Low Defect Density)**

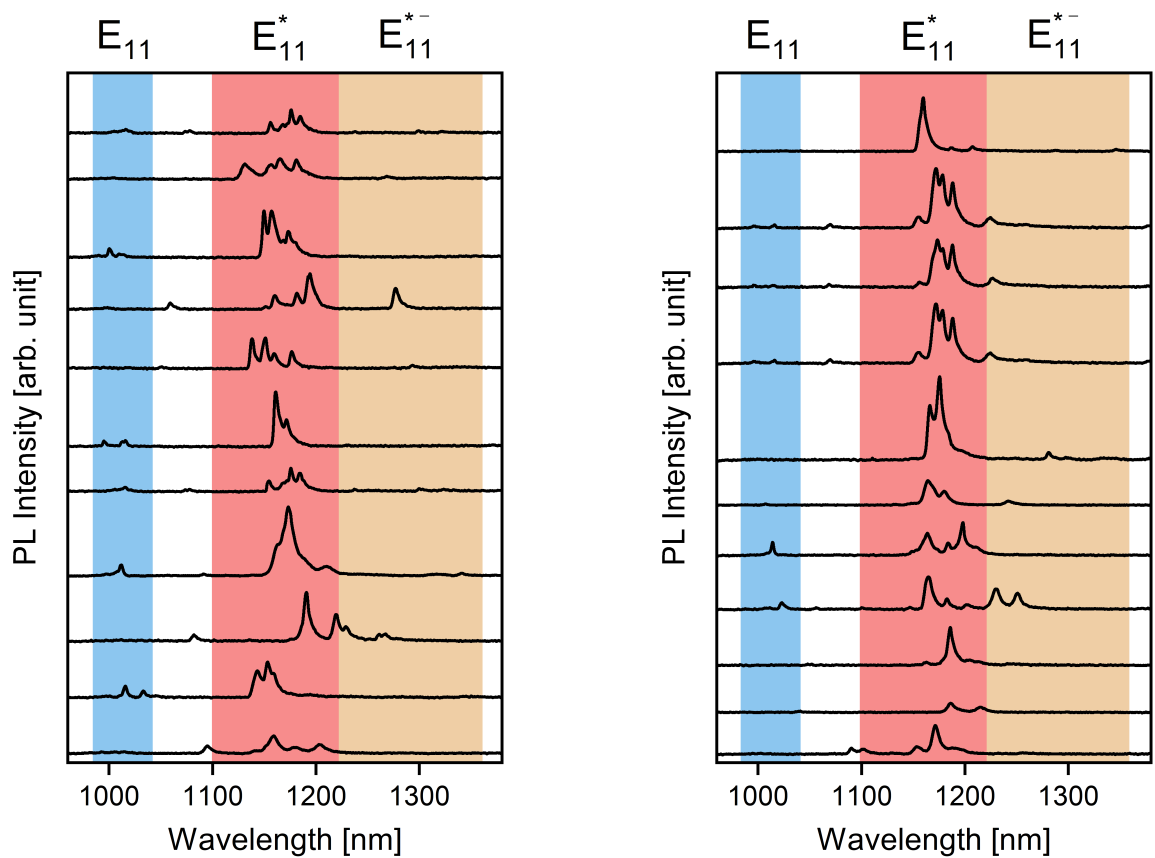

**Figure S8.** Low-temperature single-SWCNT PL spectra of (6,5) SWCNTs with a low degree of *sp*$^3$ functionalization (nominal defect density ~4 µm$^{-1}$) embedded in a polystyrene matrix, recorded at 4.6 K. The spectral ranges of the $E_{11}$ (blue) and defect emission peaks (red, orange) are highlighted.

**Low-Temperature PL Spectra of Single Nanotubes (Medium Defect Density)**

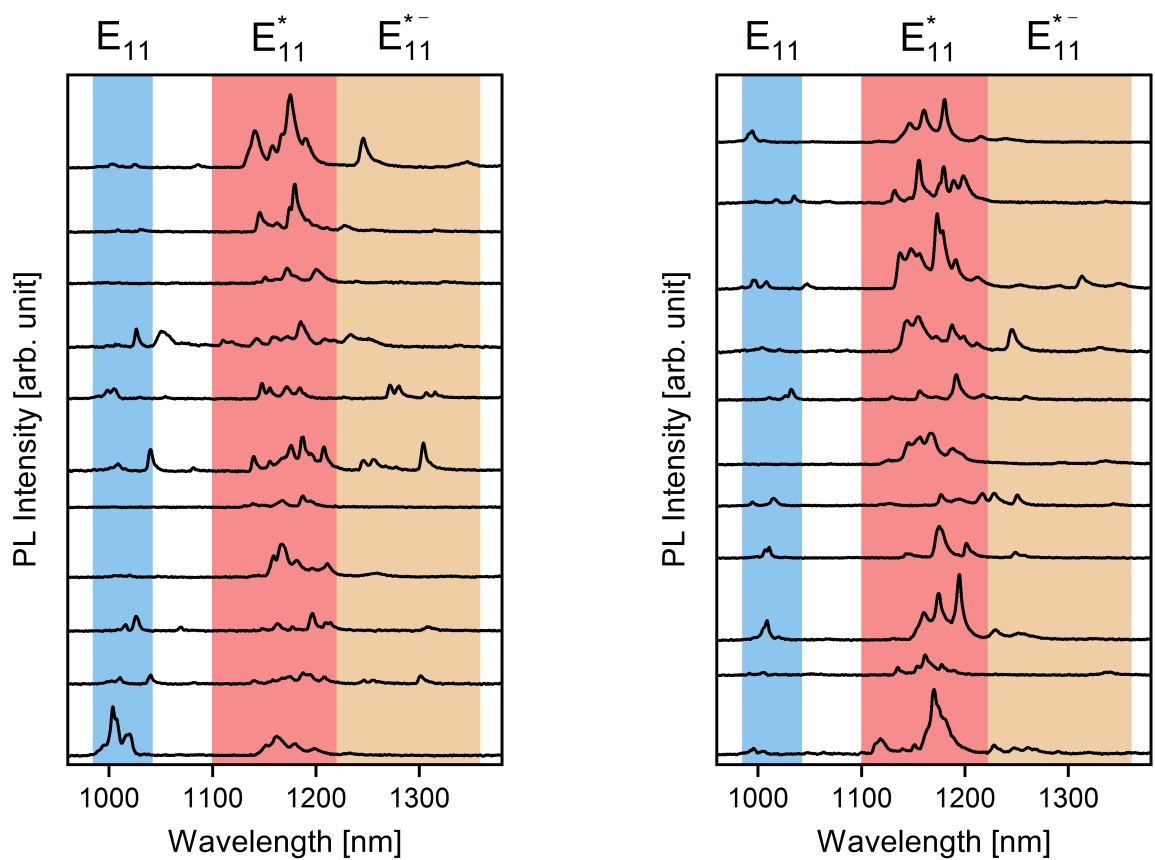

**Figure S9.** Low-temperature single-SWCNT PL spectra of (6,5) SWCNTs with a medium degree of $sp^3$ functionalization (nominal defect density ~8 μm$^{-1}$), embedded in a polystyrene matrix, recorded at 4.6 K. The spectral ranges of the $E_{11}$ (blue) and defect emission peaks (red, orange) are highlighted.

**Atomic Force Micrographs of SFM and Tip-Sonicated (6,5) SWCNTs**

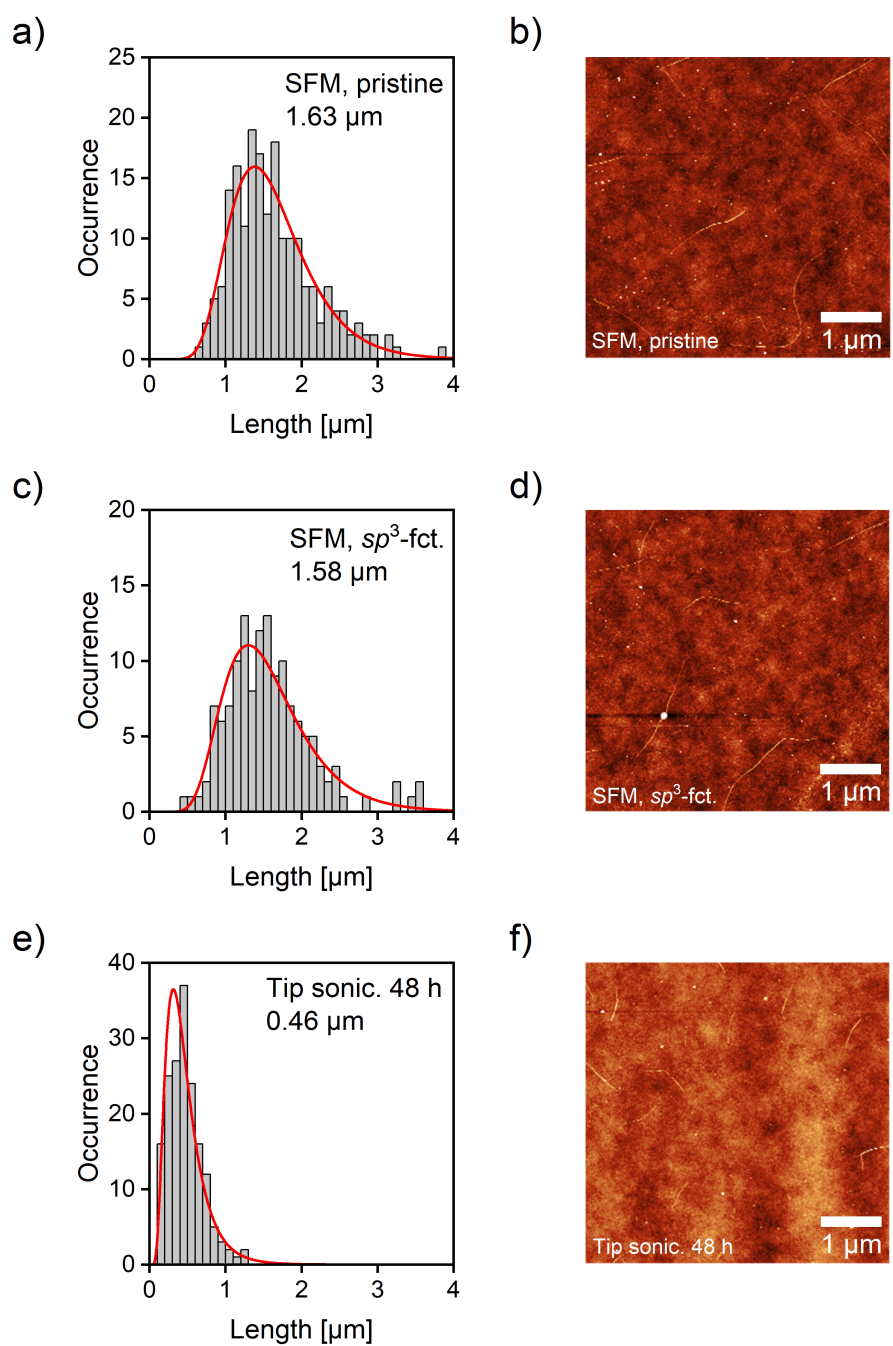

**Figure S10.** Nanotube length histograms and representative atomic force micrographs (5x5 μm$^2$) of polymer-wrapped (6,5) SWCNTs. **(a, b)** Pristine SFM nanotubes. **(c, d)** Identical batch, functionalized with 200 mg·L$^{-1}$ of DzNO$_2$. From the same dispersion, samples for low-temperature photoluminescence spectroscopy were prepared (see **Figures 4b** and **4d** in the main manuscript and **Figure S9**). **(e, f)** SFM nanotubes after 48 h of tip sonication.

**Tip-Sonicated (6,5) SWCNTs Functionalized with DzNO$_2$**

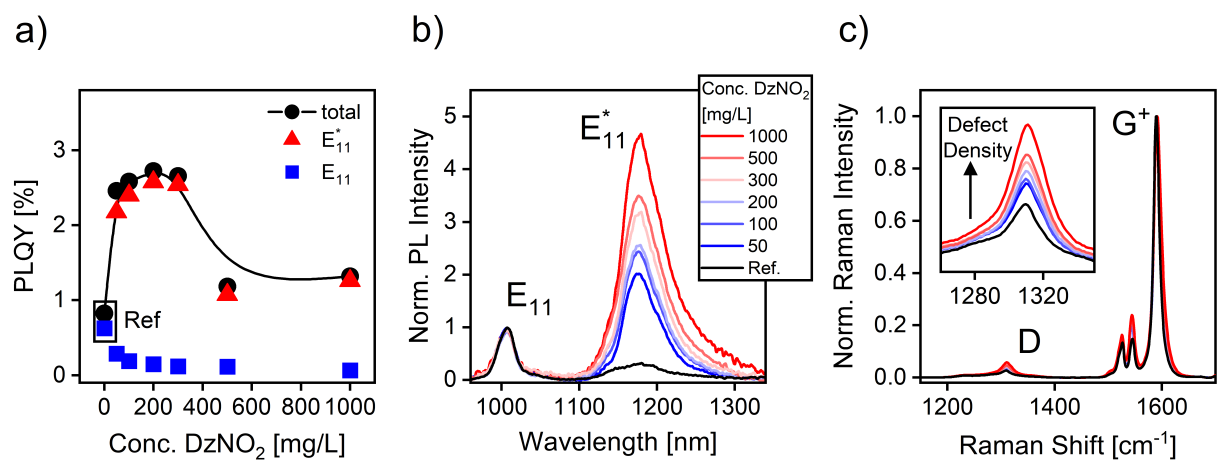

**Figure S11.** Spectroscopic data of tip-sonicated (6,5) SWCNTs wrapped with PFO-BPy and functionalized with different concentrations of DzNO$_2$. **(a)** Spectrally separated PLQY contributions of the $E_{11}$ emission and the $E_{11}$* defect emission. The solid line is a guide to the eye. **(b)** Normalized photoluminescence spectra, acquired under pulsed excitation at the $E_{22}$ transition (575 nm, ~0.025 mJ·cm$^{-2}$). **(c)** Averaged and normalized Raman spectra and zoom-in on the D mode region, recorded with a 532 nm laser at a power density of 4.1 kW·cm$^{-2}$.

**Aqueous Dispersions of (6,5) SWCNTs Functionalized with DzNO₂**

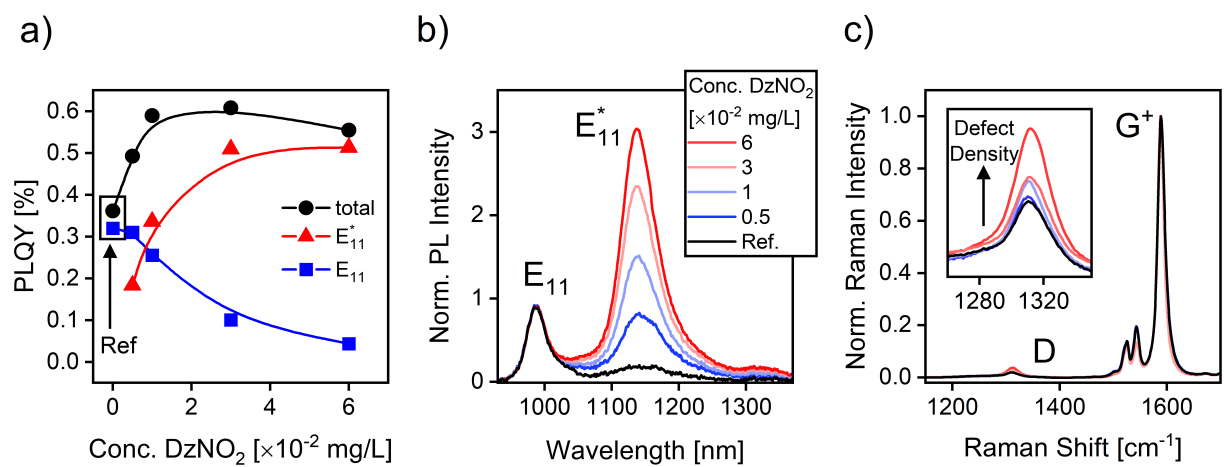

**Figure S12.** Spectroscopic data of aqueous dispersions of (6,5) SWCNTs sorted by ATPE and functionalized with different concentrations of DzNO$_2$. **(a)** Spectrally separated PLQY contributions of the $E_{11}$ emission and the $E_{11}$* defect emission. Solid lines are guides to the eye. **(b)** Normalized photoluminescence spectra, acquired under pulsed excitation at the $E_{22}$ transition (575 nm, ~0.025 mJ·cm$^{-2}$). **(c)** Averaged and normalized Raman spectra and zoom-in on the D mode region, recorded with a 532 nm laser at a power density of 4.1 kW·cm$^{-2}$.

S-20

**(7,5) SWCNTs Functionalized with DzNO₂**

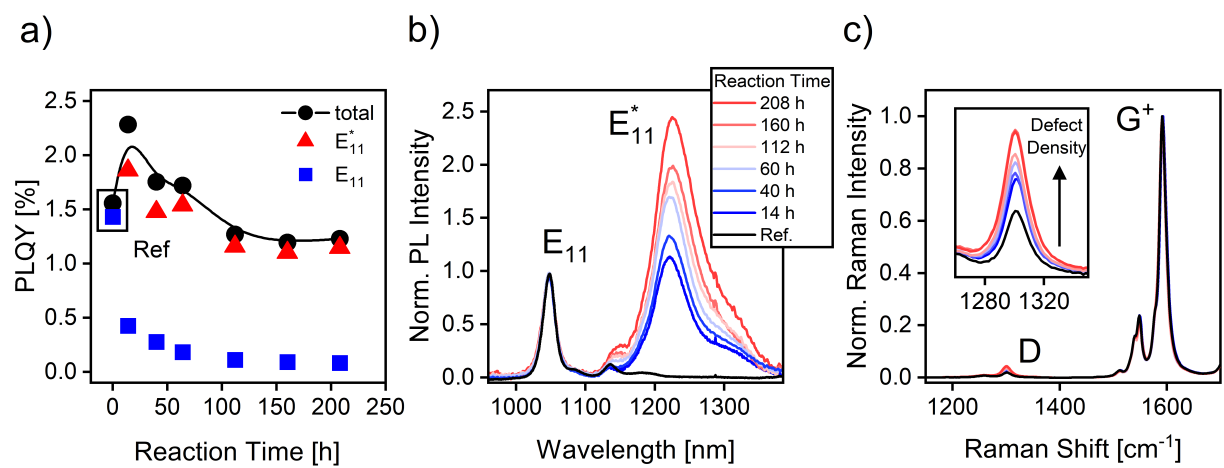

**Figure S13.** Spectroscopic data of (7,5) SWCNTs wrapped by PFO and functionalized with different concentrations of DzNO$_2$. **(a)** Spectrally separated PLQY contributions of the $E_{11}$ emission and the $E_{11}$* defect emission. The solid line is a guide to the eye. **(b)** Normalized photoluminescence spectra, acquired under pulsed excitation at the $E_{22}$ transition (652 nm, ~0.025 mJ·cm$^{-2}$). **(c)** Averaged and normalized Raman spectra and zoom-in on the D mode region, recorded with a 633 nm laser at a power density of 0.8 kW·cm$^{-2}$.

**Differential Integrated Raman D/G⁺ Ratio *versus* Average Defect Distance**

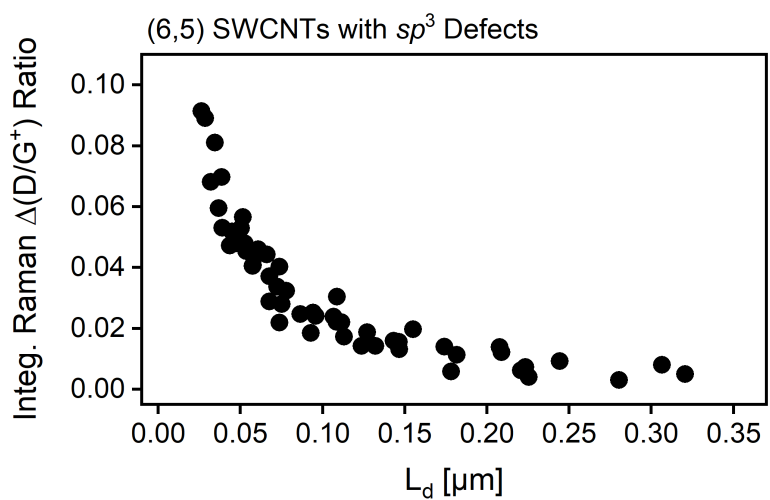

**Figure S14.** Correlation of differential integrated Raman D/G⁺ ratios and average $sp^3$ defect distances $L_d$ on functionalized (6,5) SWCNTs (including polymer-wrapped, tip-sonicated, and aqueous dispersions of SWCNTs).

**Spectroscopic Absorption Metrics of (6,5) SWCNTs with *sp*³ Defects**

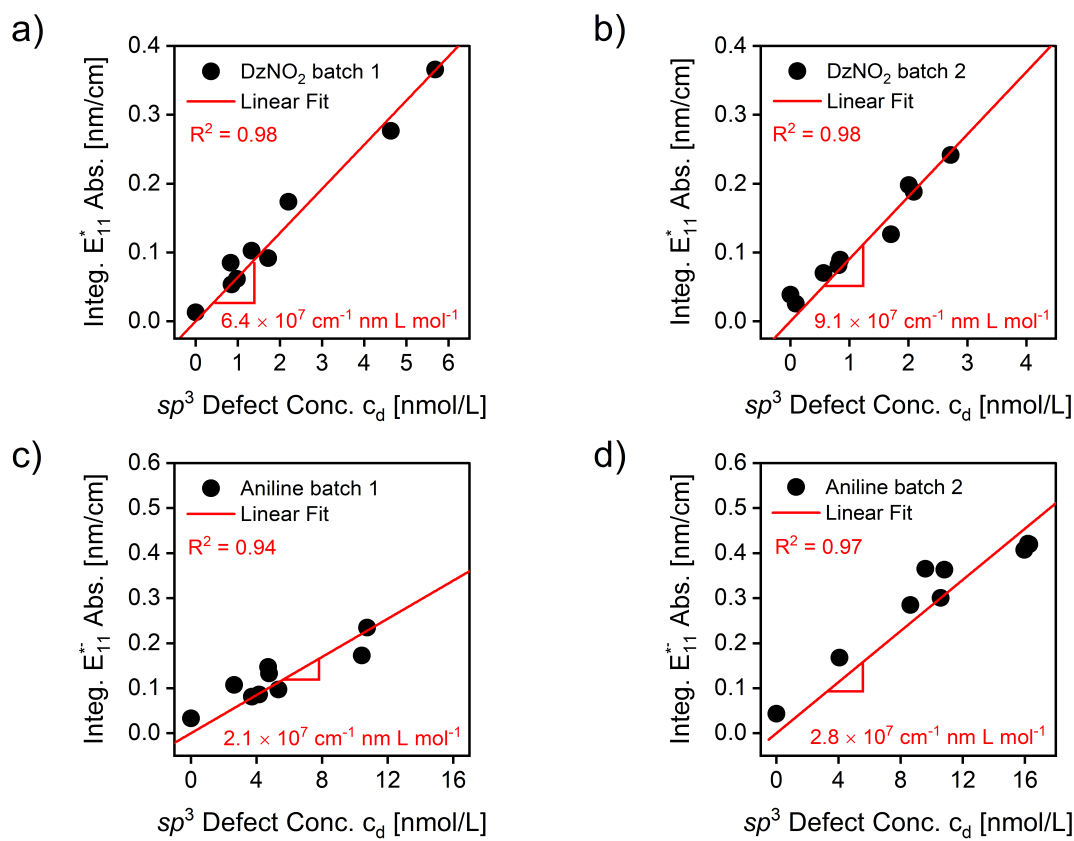

**Figure S15.** Spectrally integrated molar near-infrared absorptivities of defect states in *sp*³-functionalized (6,5) SWCNTs wrapped with PFO-BPy. Red solid lines are fits to the data. **(a, b)** Data sets for two batches of SWCNTs functionalized with $E_{11}$* defects. **(c, d)** Data sets for two batches of SWCNTs functionalized with $E_{11}$*⁻ defects.